\documentclass[aps,showpacs,showkeys,nofootinbib,superscriptaddress]{revtex4}
\usepackage{natbib}
\usepackage{amsbsy,amsmath,amsfonts,bm}
\usepackage{graphicx}
\usepackage{epsf}
\usepackage{float}
\usepackage{rotating}
\usepackage{color}
\usepackage{colordvi}
\usepackage[dvipsnames]{xcolor}
\usepackage{graphicx}
\usepackage{calrsfs} 
\usepackage{bm} 

\definecolor{green1}{rgb}{0,0,.5}
\definecolor{rosita}{rgb}{0.1,0,0}
\definecolor{miRojillo}{rgb}{1.0,0.3,0.3}

\newcommand{\bmath}{\begin{displaymath}}
\newcommand{\emath}{\end{displaymath}}
\newcommand{\beqn}{\begin{eqnarray}}
\newcommand{\eeqn}{\end{eqnarray}}
\newcommand{\beqns}{\begin{eqnarray*}}
\newcommand{\eeqns}{\end{eqnarray*}}

\newcommand{\MeV}{{\rm \,MeV}}
\newcommand{\tr}{{\rm tr }}

\newcommand{\HQSS}{{\rm HQSS} }
\newcommand{\WT}{{\rm WT}}
\newcommand{\SU}{\mbox{SU}}

\newcommand{\pv}{i\overset{\leftrightarrow}{\partial}\!{}_v}

\newcommand{\ben}{\begin{enumerate}}
\newcommand{\een}{\end{enumerate}}

\newcommand{\be}{\begin{equation}}
\newcommand{\ee}{\end{equation}}
\newcommand{\bea}{\begin{eqnarray}}
\newcommand{\eea}{\end{eqnarray}}
\newcommand{\ds}{\begin{displaystyle}}
\renewcommand{\ss}{\begin{scriptstyle}}

\newcommand{\ignore}[1]{}

\newcommand{\ba}{\begin{eqnarray}}
\newcommand{\ea}{\end{eqnarray}}

\newcommand{\Db}{{{\bar{D}} }}

\begin{document}

\markboth{C. García-Recio, D. Gamermann, J. Nieves, O. Romanets, L.L. Salcedo
  and L. Tolós}
{Charming Baryons}

%

\title{CHARMING BARYONS}

\author{C. GARCIA-RECIO\footnote{ Invited talk at NSTAR 2013}
 ~and L. L. SALCEDO}
\affiliation{Departamento~de~F{\'\i}sica~At\'omica, Molecular~y~Nuclear, and
  Instituto Carlos I de F{\'\i}sica Te\'orica y Computacional,
  Universidad~de~Granada, E-18071~Granada, Spain
}
\author{D. GAMERMANN}
\affiliation{
Universidad
  Cat\'olica de Valencia San Vicente M\'artir, \\ Guillem de Castro 94,
  E-46003, Valencia, Spain 
}
\author{J. NIEVES}
\affiliation{Instituto~de~F{\'\i}sica~Corpuscular~(centro~mixto~CSIC-UV),\\
  Institutos~de~Investigaci\'on~de~Paterna, Aptdo.~22085,~46071,~Valencia,
  Spain,
}
\author{O. ROMANETS}
\affiliation{ KVI,
  University~of~Groningen, Zernikelaan~25,~9747AA~Groningen, The~Netherlands
 }
\author{L. TOLOS}
\affiliation{ Institut~de~Ci\`encies~de~l'Espai~(IEEC/CSIC),
  Campus~Universitat~Aut\`onoma~de~Barcelona, Facultat~de~Ci\`encies,
  Torre~C5,~E-08193~Bellaterra,~Spain \\
Frankfurt Institute for Advanced Studies, \\
Johann Wolfgang Goethe University, Ruth-Moufang-Str.~1,
60438 Frankfurt am Main, Germany
}

\begin{abstract}
We study odd-parity baryonic resonances with one heavy and three light
flavors, dynamically generated by meson-baryon interactions. Special attention
is paid to Heavy Quark Spin Symmetry ($\HQSS$), hence pseudoscalar and vector
mesons and baryons with $J^\pi=1/2^+$ and $3/2^+$ are considered as
constituent hadrons. For the hidden-charm sector ($N_c=N_{\bar{c}}=1$), the
meson-baryon Lagrangian with Heavy Flavor Symmetry is constructed by a {\it
  minimal} extension of the SU(3) Weinberg-Tomozawa (WT) Lagrangian to fulfill
\HQSS, such that not new parameters are needed.  This interaction can be
presented in different formal ways: as a Field Lagrangian, as Hadron
creation-annihilation operators, as SU(6)$\times$\HQSS group projectors and as
multichannel matrices. The multichannel Bethe-Salpeter equation is solved for
odd-parity light baryons, hidden-charm $N$ and $\Delta$ and Beauty Baryons
($\Lambda_b$). Results of calculations with this model are shown in comparison
with other models and experimental values for baryonic resonances.
\end{abstract}



\maketitle

\section{Introduction}

The study of baryonic resonances with charm or beauty is an active field of
research, stimulated by current experiments, such as CLEO, Belle and BABAR, as
well as planned ones, in PANDA and CBM at FAIR. In this regard
hot topics are to find exotic states and to uncover the nature of such
resonances, either as most quark model-like or as hadronic molecule-like.

In the molecular description, the baryonic resonance is interpreted as a
baryon-meson system. Theoretical work in this direction relies on unitarized
coupled-channels dynamics with various
kernels\cite{Tolos:2004yg,Tolos:2005ft, 
Lutz:2003jw,Hofmann:2005sw, Hofmann:2006qx, 
Lutz:2005vx,Mizutani:2006vq,Lutz:2005ip,Tolos:2007vh,JimenezTejero:2009vq,
Haidenbauer:2007jq,Haidenbauer:2010ch,Wu:2010jy,
Wu:2010vk,Wu:2012md,Oset:2012ap}. The same approach has been applied
previously in the strangeness
sector\cite{Oset:1997it, 
Kaiser:1995eg,Oller:2000fj,Kaiser:1996js,Lutz:2001yb,Kaiser:1995cy,
Inoue:2001ip,Oset:2001cn,Borasoy:2005ie,Kolomeitsev:2003kt, 
GarciaRecio:2002td,Sarkar:2004jh,Hyodo:2008xr,
Nieves:2001wt}. In
our own approach to the problem we follow this route and use a contact
interaction based on extending the Weinberg-Tomozawa term to enjoy SU(6)
spin-flavor invariance and to comply with heavy-quark spin
symmetry\cite{Isgur:1989vq,Neubert:1993mb,Manohar:2000dt}. In particular, the
latter requirement is unavoidable as HQSS becomes a rather good symmetry
already at the charm scale and more so at the bottom scale, and explains
noticeable regularities in the properties of the low-lying baryons and mesons
with heavy quarks. The enforcement of HQSS has not always been applied in
previous approaches and it is one of the strong points of our treatment. Our
model applies to the study of odd-parity baryonic states with open or hidden-charm and with or without strangeness and naturally extends to bottom states.

\section{The model}

The standard WT interaction takes the form
\begin{equation}
V_\WT = \frac{K(s)}{2f^2} J_P^i J_T^i,
\qquad
i=1,\ldots,N_F^2-1 ,
\label{eq:2.1}
\end{equation}
where $J_{P,T}^i$ denote the $\SU(N_F)$ flavor generators for projectile (a
pseudo-Goldstone boson) and target (a baryon), with $N_F=3$, and near
threshold $K(s)$ is fixed by the chiral dynamics. Its spin-flavor extension is
immediate by using the corresponding $\SU(2N_F)$
generators\cite{GarciaRecio:2005hy,GarciaRecio:2006wb}. Since spin-flavor
multiplets are formed by $0^-$ and $1^-$ states in the meson sector and
$1/2^+$ and $3/2^+$ states in the baryon sector, all these low-lying states
are included in the coupled-channels dynamics. For the three flavors, this
involves the pion octet and the $\rho$ nonet, and the nucleon octet and the
$\Delta$ decuplet. Results in this sector have been reported
in Ref.~\cite{Gamermann:2011mq}.

In order to include charm, we first extend the model to $N_F=4$ and then break
$\SU(4)$ flavor (and $\SU(8)$ spin-flavor) explicitly in order to enforce
HQSS. Specifically, the SU(8)-extended meson-baryon WT interaction
\begin{equation}
{\mathcal H}_\WT^{\rm sf}(x)
= -\frac{{\rm i}}{4f^2} :[\Phi, \partial_0 \Phi]^A{}_B
{\cal B}^\dagger_{ACD} {\cal B}^{BCD}:
,
\quad
A,B,\ldots = 1,\ldots,2N_F
\label{eq:2.9}
\end{equation}
contains two types of contributions, when analyzed in terms of
quarks\cite{Garcia-Recio:2013gaa}. The first type of contributions corresponds
to pure exchange of quarks/antiquarks between meson and baryon, while the
second type corresponds to the antiquark annihilation with a quark of the
baryon (followed by antiquark-quark creation). Heavy quark-antiquark
annihilation violates HQSS and such terms are suppressed in QCD by a factor
proportional to the inverse of the heavy quark mass. For simplicity we
completely remove those terms, thus guaranteeing exact HQSS of our model. The
interaction so obtained enjoys chiral symmetry and full $\SU(6)\times \HQSS$
invariance, however the latter invariance is actually broken in our
calculations by using the physical values for the hadron masses and the meson
decay constants. The model contains no free parameters. The only freedom lies
in the choice of the baryon-meson loop renormalization, for which we apply the
prescription in Refs.~\cite{Hofmann:2005sw,Hofmann:2006qx}.

An analysis of the allowed interactions consistent with flavor SU(3) and HQSS
in the hidden-charm sector (meson-baryon states with exactly one charm quark
and one charm antiquark) shows that twelve independent interactions are
allowed, corresponding to the Lagrangians\cite{Garcia-Recio:2013gaa}
\begin{eqnarray}
\mathcal{L}_1(x) &=&
g_1\, \overline{\Sigma}{}^a{}_b
\Sigma^b{}_a
\,\tr(\overline{\bm{\psi}}\,\pv\bm{\psi}) 
, \label{eq:g1}
\\
\mathcal{L}_2(x) &=&
g_2\, \frac{1}{3!}\overline{\Delta}{}_{abc}^\mu \Delta^{abc}_\mu 
\,\tr( \overline{\bm{\psi}} \,\pv \bm{\psi} )
,
\\
\mathcal{L}_3(x) &=&
g_3\, \overline{\bm{\Xi}}{}_{\bm{c}}^a \bm{\psi}(-\pv)
\overline{\bm{\bar{D}}}_b \Sigma^b{}_a 
+\mbox{h.c.}
,
\\
\mathcal{L}_4(x) &=&
g_4\, \epsilon^{bcd}
\overline{\bm{\Sigma}}{}^\mu_{\bm{c}}{}_{ab}\bm{\psi}(-\pv)
\overline{\bm{\bar{D}}}_c \gamma_\mu\gamma_5 \Sigma^a{}_d
+\mbox{h.c.}
,
\\
\mathcal{L}_5(x) &=&
g_5\, \frac{1}{2}
\overline{\bm{\Sigma}}{}^\mu_{\bm{c}}{}_{ab}
\bm{\psi} (-\pv)
\overline{\bm{\bar{D}}}_c
\,\Delta^{abc}_\mu 
+\mbox{h.c.}
,
\\
\mathcal{L}_6(x) &=&
g_6\, \overline{\bm{\Xi}}{}_{\bm{c}}^a \bm{\Xi}_{\bm{c}}{}_a 
\,\tr(\overline{\bm{\bar{D}}}_b \,\pv \bm{\bar{D}}^b) 
,
\\
\mathcal{L}_7(x) &=&
g_7 \,\overline{\bm{\Xi}}{}_{\bm{c}}^a \bm{\Xi}_{\bm{c}}{}_b 
\,\tr(\overline{\bm{\bar{D}}}_a \,\pv \bm{\bar{D}}^b) \label{eq:eqg7}
,
\\
\mathcal{L}_8(x) &=&
g_8 \,
 \epsilon^{bcd}
\overline{\bm{\Sigma}}{}^\mu_{\bm{c}}{}_{ab}
\bm{\Xi}_{\bm{c}}{}_d
\,\tr(
\overline{\bm{\bar{D}}}_c
\gamma_\mu\gamma_5
\,\pv
\bm{\bar{D}}^a
)
+\mbox{h.c.}
,
\\
\mathcal{L}_9(x)+\mathcal{L}_{10}(x) &=&
\frac{1}{2}
\overline{\bm{\Sigma}}{}^\mu_{\bm{c}}{}_{ab}
\bm{\Sigma}_{\bm{c}}^\nu{}^{ab}
\,\tr(
\overline{\bm{\bar{D}}}_c
 (g_9 \,g_{\mu\nu} + g_{10} \,i\sigma_{\mu\nu})
\,\pv \bm{\bar{D}}^c
)
,
\\
\mathcal{L}_{11}(x)+\mathcal{L}_{12}(x) &=&
\overline{\bm{\Sigma}}{}^\mu_{\bm{c}}{}_{ac}
\bm{\Sigma}_{\bm{c}}^\nu{}^{bc}
\,\tr(
\overline{\bm{\bar{D}}}_b
 (g_{11} \,g_{\mu\nu} + g_{12} \,i\sigma_{\mu\nu})
\,\pv \bm{\bar{D}}^a
)
. \label{eq:g11}
\end{eqnarray}
Our model gives the following values for the parameters (where we have defined $\hat{g}_i = 4f^2 g_i$):
\begin{eqnarray}
&&
\hat{g}_1 = 0 ,\quad
\hat{g}_2 = 0 ,\quad
\hat{g}_3 = \sqrt{\frac{3}{2}} ,\quad
\hat{g}_4 =  \sqrt{\frac{1}{6}} ,\quad
\hat{g}_5 =  -1 ,\quad
\hat{g}_6 =  \frac{1}{2} ,
\nonumber \\
&&
\hat{g}_7 = -\frac{1}{2} ,\quad
\hat{g}_8 = \frac{1}{2} ,\quad
\hat{g}_9 = 0 ,\quad
\hat{g}_{10} =  0 ,\quad
\hat{g}_{11} =  -\frac{1}{2} ,\quad
\hat{g}_{12} = -\frac{1}{2} \label{eq:gs}
.
\end{eqnarray}

\section{Selected results}

The extended WT model just described (as well as its bottomed version) has
been applied to a number of
cases\cite{Garcia-Recio:2013gaa,GarciaRecio:2008dp,Gamermann:2010zz%
  ,Romanets:2012hm,GarciaRecio:2012db}. Here we touch just three topics.

For the hidden-charm sector, we show results in Table~\ref{tab1} for $N$-like
and $\Delta$-like baryons. The widths are either zero or small in all
cases. The dominant channels in each case are also displayed. The group labels
are assigned by adiabatic breaking of the symmetries. The states are arranged
into HQSS multiplets. Some of these states are also found within other
approaches\cite{Hofmann:2005sw,Hofmann:2006qx%
  ,Wu:2010vk,Wu:2012md,Yuan:2012wz} with various values for the mass of the
resonances. However, at least for the hidden gauge approach, it has been shown
in Ref.~\cite{Xiao:2013yca} that the shifts in mass reflect more the different
use of renormalization prescriptions than differences in the interactions
themselves.

\begin{figure}[ph]
\centerline{ 
\epsfxsize = 80 mm  \epsfbox{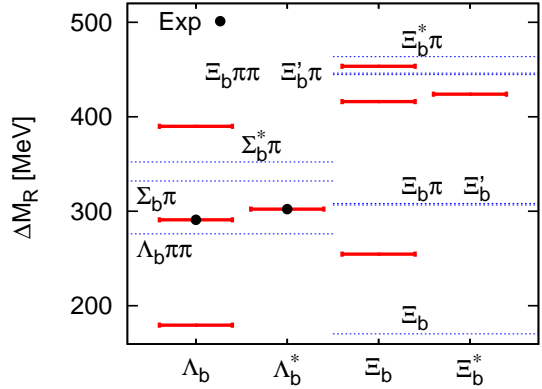}
      }

\caption{ Predicted bottomed baryonic states\cite{GarciaRecio:2012db}. Also shown are
  the experimentally observed $\Lambda_b^0(5912)$ and $\Lambda_b^0(5920)$
  states\cite{Aaij:2012da} and some relevant hadronic
  thresholds. $\Delta\,M_R=M_R-M_{\Lambda_b(g.s)}$. }
\label{f1}
\end{figure}

\begin{table*}[ph]
  \caption{Odd-parity hidden-charm $N$ and $\Delta$ resonances found within the
    model, with their group labels, mass, dominant couplings, isospin and
    spin.  Resonances with equal $\SU(6)\times\HQSS$ and $\SU(3)\times\HQSS$
    labels form HQSS multiplets, and they are collected in consecutive rows.  }
{\begin{tabular}{ c  c  c  c  c  c c }\toprule
 $\SU(6)\times\HQSS$  & $\SU(3)\times\HQSS$    & & Couplings &  \\
 irrep & irrep & $M_R [\MeV]$   &  to main channels & $I$ & $J$ \\ 
\hline
 $\bf{70_{2,0}}$  & $\bf{(8_2)_{2,0}}$  &  3918.3  & 
$g_{\Lambda_c \bar{D}}=3.1$, 
$g_{\Sigma_c \bar{D}^*}=2.6$, 
$g_{\Sigma_c^* \bar{D}^*}=2.6$
& $1/2$ & $1/2$ 
\\
 $\bf{70_{2,0}}$ &   $\bf{(8_2)_{2,0}}$  & 3926.0 & 
$g_{\Lambda_c \bar{D}^*}=3.0$, 
$g_{\Sigma_c \bar{D}}=4.2$,  
& $1/2$  & $1/2$ \\
$\bf{70_{2,0}}$ & $\bf{(8_2)_{2,0}}$ &  3946.1  & 
$g_{\Lambda_c \bar{D}^*}=3.4$,  
$g_{\Sigma_c^* \bar{D}}=3.6$,  
& $1/2$ & $3/2$ \\

    $\bf{70_{2,0}}$  &  $\bf{(8_4)_{2,0}}$  & 3974.3 & 
 $g_{\Sigma_c \bar{D}^*}=3.4$, $g_{\Sigma_c^* \bar{D}^*}=3.1$
 & $1/2$  & $1/2$ \\

    $\bf{70_{2,0}}$  &  $\bf{(8_4)_{2,0}}$ &   3986.5    & 
  $g_{\Sigma_c^* \bar{D}}=2.7$,   $g_{\Sigma_c \bar{D}^*}=4.3$,
& $1/2$ & $3/2$ \\

    $\bf{70_{2,0}}$  &  $\bf{(8_4)_{2,0}}$ &  4005.8 &    
$g_{\Sigma_c \bar{D}^*}=3.2$,
$g_{\Sigma_c^* \bar{D}^*}=4.2$
& $1/2$ & $3/2$ \\

$\bf{70_{2,0}}$  &  $\bf{(8_4)_{2,0}}$ & 4027.1  &   $g_{\Sigma_c^*~\bar{D}^*}=5.6$
& $1/2$ & $5/2$ \\
\hline
$\bf{70_{2,0}}$ &  $\bf{(10_2)_{2,0}}$  &4005.8   & 
$g_{\Sigma_c \bar{D}}=2.7$, $g_{\Sigma_c \bar{D}^*}=4.4$, 
& $3/2$ & 1/2 \\
$\bf{70_{2,0}}$ &  $\bf{(10_2)_{2,0}}$   &  4032.5   &  
$g_{\Sigma_c^* \bar{D}}=2.9$,  
$g_{\Sigma_c^* \bar{D}^*}=4.1$  & $3/2$ &  3/2  \\
 $\bf{70_{2,0}}$ & $\bf{(10_2)_{2,0}}$  & 4050.0  &  
 $g_{\Sigma_c \bar{D}^*}=1.9$,  $g_{\Sigma_c^* \bar{D}^*}=5.1$
 & $3/2$ & $1/2$ \\
$\bf{56_{2,0}}$ & $\bf{(10_4)_{2,0}}$  & 4306.2 (cusp) & $g_{\Delta J/\psi}=1.3$,
 & $3/2$ & $1/2$ \\
$\bf{56_{2,0}}$ & $\bf{(10_4)_{2,0}}$ &  4306.8 (cusp)  & 
$g_{\Delta J/\psi}=0.8$,     & $3/2$ &  $3/2$  \\
\hline
\end{tabular}
\label{tab1}}
\end{table*}

The negative charm baryons are exotic as they are necessarily
pentaquarks rather than three-quark states. In this sector several states are
obtained within our model\cite{Gamermann:2010zz}. Of particular interest is
the isoscalar strangeless $1/2^-$ baryon at $2805\MeV$. This is a bound state
of $\Db N$ with a large component of $\Db^* N$ (this state is part of
a HQSS doublet). The binding energy is just $1\MeV$, and the same prediction has
been obtained in\cite{Yasui:2009bz}, also by enforcing HQSS. This state plays
an important role in the dynamics of $\Db$-nucleus systems, including $D^-$
mesic atoms and $\Db^0$ mesic nuclei. As discussed
in Ref.~\cite{GarciaRecio:2011xt}, $\Db N$ is the lightest hadronic channel for
those quantum numbers. This means that, after Coulomb and particle-hole
cascading, a $D^-$ would remain inside the nucleus, perhaps forming a
pentaquark, awaiting for weak decay.

Finally, we present results for $1/2^-$ and $3/2^-$ $\Lambda_b$ baryons. Such
states have recently been found by the LHCb Collaboration\cite{Aaij:2012da}
with masses remarkably close to those predicted by the relativized quark
model\cite{Capstick:1986bm}. The same states are dynamically generated within
our model\cite{GarciaRecio:2012db} as part of a full $\SU(3)\times\HQSS$
multiplet which also involves $\Xi_b$ states with $J=1/2$ and $3/2$, not yet
observed. This is displayed in Fig. \ref{f1}, together with relevant hadronic
thresholds. We find a close analogy with the strange and charm sectors. In
particular, $\Lambda_b(5912)$ is part of a two-pole structure similar to that
present in the $\Lambda(1405)$ and $\Lambda_c(2995)$ resonances\cite{Magas:2005vu}.

\section{Summary}

We have reviewed a model for dynamically generated odd-parity
baryon resonances with charm or bottom, through coupled-channels
unitarization. The model is a minimal one based on chiral, spin-flavor and
heavy-quark spin invariances and it has no free parameters in the interaction.
Results have been reported in several sectors.

\section*{Acknowledgments}
Supported by Spanish Ministerio de Econom\'{\i}a y Competitividad
(FIS2011-28853-C02-02, FIS2011-24149, FPA2010-16963), Junta de Andaluc\'{\i}a
(FQM-225), Generalitat Valenciana (PROMETEO/2009/0090) and EU HadronPhysics2
project (grant 227431). O. R. acknowledges support from the Rosalind Franklin
Fellowship. L. T. acknowledges support from RyC Program, and
FP7-PEOPLE-2011-CIG (PCIG09-GA-2011-291679).


\end{document}